\documentclass[a4paper, 12pt, openany, oneside]{article}
\usepackage[cp1251]{inputenc}
\usepackage[english]{babel}
\usepackage{ graphics,amsfonts, amsmath,bm,amssymb, array, eucal}
\usepackage[dvips]{graphicx}
\makeatletter
\renewcommand{\@biblabel}[1]{#1.\hfill}

\newcommand{\intl}{\int\limits}

\renewcommand{\Re}{\mathop{\rm Re\,}}

\newcommand{\Res}{\mathop{\rm Res\,}}
\newcommand{\sign}{\mathop{\rm sign\,}}
\makeatother {
\usepackage{indentfirst}

\begin{document}
\thispagestyle{empty} \large

\renewcommand{\refname}{\normalsize\begin{center}\bf
REFERENCES\end{center}}

\begin{center}
{\bf LANDAU'S HALF-SPACE PROBLEM OF DEGENERATE PLASMA OSCILLATIONS 
WITH SPECULAR -- ACCOMMODATIVE BOUNDARY CONDITIONS}
\end{center}

\begin{center}
  \bf  N.V. Gritsienko\footnote{$natafmf@yandex.ru$}, 
A. V. Latyshev\footnote{$avlatyshev@mail.ru$} and
  A. A. Yushkanov\footnote{$yushkanov@inbox.ru$}
\end{center}\medskip

\begin{center}
{\it Faculty of Physics and Mathematics,\\ Moscow State Regional
University,  105005,\\ Moscow, Radio st., 10--A}
\end{center}\medskip

\renewcommand{\abstractname}{\ }
\newcommand{\mc}[1]{\mathcal{#1}}
\newcommand{\E}{\mc{E}}

\thispagestyle{empty} \large

\date{\today}

\begin{abstract}
In the present paper the linearized problem of half-space 
plasma oscillations
in external longitudinal alternating electric field is solved
analytically. Specular -- accommoda\-ti\-ve boundary conditions of
electron reflection from the plasma boundary are considered.
Coefficients of continuous and discrete spectra of the problem are
found, and electron distribution function on the plasma boundary and
electric field are expressed in explicit form.

Refs. 34. Figs. 2.\\

\end{abstract}


\begin{center}\bf
1. INTRODUCTION
\end{center}

The present paper is devoted to degenerate electron plasma behaviour
research. Analysis of processes taking place in plasma under effect
of external electric field, plasma waves oscillations with various
types of condi\-tions of electron reflection from the boundary has
important significance today in connection with problems of such
intensively developing fields as microelec\-tronics and nanotechnologies
\cite{1} -- \cite{Liboff}.

The concept of "plasma"\, appeared in the works of Tonks and Langmuir
for the first time (see \cite{Langmuir}--\cite{Tonks2}),
the concept of "plasma frequency"\, was introduced in the same works
and first questions of plasma oscillations were considered there.
However, in these works equation for the electric field was
considered separately from the kinetic equation.

A.A. Vlasov \cite{Vlasov} for the first time introduced the concept
of "self-consistent electric field"\, and added the corresponding item
to the kinetic equation. Now equations describing plasma behaviour
consist of anchor system of equations of Maxwell and Boltzmann.
The problem of electron plasma oscillations was considered by A.A.
Vlasov \cite{Vlasov} by means of solution of the kinetic equation
which included self-consistent electric field.

L.D. Landau \cite{Landau} had supposed that outside of the half-space
containing degenerate plasma external electromagnetic field causing
oscillations in plasma is situated. By this Landau has formulated a
boundary condition on the plasma boundary. After that the problem
of plasma oscillation turns out to be formulated correctly as a
boundary problem of mathematical physics.

In \cite{Landau}  L.D. Landau has solved analytically by Fourier
series the problem
of collisionless plasma behaviour in a half-space, situated in
external lon\-gi\-tu\-dinal (perpen\-di\-cular to the surface) electric
field, in conditions of specular reflection of electrons
from the boundary.

Further the problem of electron plasma oscillations was considered
by many authors. Full analytical solution of the problem is given
in the works \cite{8} and \cite{6}.

This problem has important significance in the theory of plasma (see,
for instance, \cite{2}, \cite{3} and the references in these works,
and also
\cite{M1}, \cite{M2}).

The problem of plasma oscillations with diffuse boundary condition
was considered in the works
\cite{Keller}, \cite{Kliewer} by method of integral transformations.
In the works \cite{Gohfeld1}, \cite{Gohfeld2} general asymptotic
analysis of electric field behaviour at the large distance from
the surface was carried out. In the work \cite{Gohfeld1} particular
significance of plasma behaviour analysis close to plasma re\-so\-nan\-ce
was shown. And in the same work \cite{Gohfeld1}  it was stated that
plasma behaviour in this case for conditions of specular and diffuse
electron scat\-te\-ring on the surface differs substantially.

In the works \cite{4} and \cite{5} general questions of this problem
solvability were considered, but diffuse boundary conditions were
taken into account. In the work \cite{5} structure of discrete
spectrum in dependence of parameters of the problem was analyzed.
The detailed analysis of the solution in general case in the works
mentioned above hasn't been carried out conside\-ring the complex
character of this solution.

The present work is a continuation of electron plasma behaviour in
external longitudinal alternating electric field research
\cite{4} -- \cite{11}.

In the present paper the linearized problem of half-space 
plasma oscillations
in external alternating electric field is solved analytically.
Specular -- accommo\-da\-ti\-ve boundary conditions for electron
reflection from the boundary are considered. In \cite{9}--\cite{11}
diffuse boundary conditions were considered.

The coefficients of continuous and discrete spectra of the problem
are obtained in the present work, which allows us to derive
expressions for electron distribution function at the boundary
of conductive medium and electric field in explicit form, to reveal
the dependence of this expressions on normal momentum accommodation
coefficient and to show that in the case when normal electron
momentum accommodation coefficient equals to zero electron
distribution function and electric field are expressed by
known formulas obtained earlier in \cite{8}, \cite{6}.

The present work is a continuation of our work \cite{Gritsienko},
in which questions of plasma waves reflection from the plane boundary
bounding degenerate plasma were considered.

Let us note, that questions of plasma oscillations are also
considered in nonlinear statement (see, for instance, the work
\cite{Stenflo},
\cite{Stenflo2}).

\begin{center}\bf
2. PROBLEM STATEMENT
\end{center}

Let degenerate plasma occupy a half-space $x>0$.

We take system of equations describing plasma behaviour.
As a  kinetic equation we take Boltzmann --- Vlasov $\tau$--model 
kinetic equation:
$$
\dfrac{\partial f}{\partial t}+\mathbf{v}\dfrac{\partial f}{\partial
\mathbf{r}}+e\mathbf{E}\dfrac{\partial f}{\partial \mathbf{p}}=
\dfrac{f_{eq}(\mathbf{r},t)-f(\mathbf{r},\mathbf{v},t)}{\tau}.
\eqno{(1.1)}
$$

Here $f=f(\mathbf{r},\mathbf{v},t)$ is the electron distribution
function, $e$ is the electron charge, $\mathbf{p}=m\mathbf{v}$
is the momentum of an electron, $m$ is the electron mass, $\tau$
is the character time between two collisions,
$\mathbf{E}=\mathbf{E}(\mathbf{r},t)$ is the self-consistent
electric field inside plasma,
$f_{eq}=f_{eq}(\mathbf{r},t)$ is the local equilibrium Fermi --- Dirac distribution function,
$ f_{eq}=\Theta(\E_F(t,x)-\E), $ where $\Theta(x)$ is the function
of Heaviside,
$$
\Theta(x)=\left\{\begin{array}{c}
                   1, \quad x>0, \\
                   0,\quad x<0,
                 \end{array}\right.
$$
$\E_F(t,x)=\frac{1}{2}mv_F^2(t,x)$ is the disturbed kinetic energy
of Fermi, $\E=\frac{1}{2}mv^2$ is the kinetic energy of electron.

Let us take the Maxwell equation for electric field
$$
{\rm div}\,{\mathbf{E}(\mathbf{r},t)}= 4\pi\rho(\mathbf{r},t).
\eqno{(1.2)}
$$
Here $\rho(\mathbf{r},t)$ is the charge density,
$$
\rho(\mathbf{r},t)=e\int (f(\mathbf{r},
\mathbf{v},t)-f_{0}(\mathbf{v})) \,d\Omega_F, \eqno{(1.3)}
$$
where
$$
d\Omega_F=\dfrac{2d^3p}{(2\pi\hbar)^3}, \qquad
d^3p=dp_xdp_ydp_z.
$$

Here $f_{0}$  is the undisturbed Fermi --- Dirac electron 
distribution function,
$$
f_{0}(\E)=\Theta(\E_F-\E),
$$
$\hbar$ is the Planck's constant, $\nu$ is the effective frequency
of electron collisions, $\nu=1/\tau$, $\E_F=\frac{1}{2}mv_F^2$ is
the undisturbed kinetic energy of Fermi, $v_F$ is the electron
velocity at the Fermi surface, which is supposed to be spherical.

We assume that external electric field outside the plasma is
perpendi\-cu\-lar to the plasma boundary and changes according to
the following law: $E_{0}\exp(-i\omega t).$

Then one can consider that self-consistent
electric field
$\mathbf{E}(\mathbf{r},t)$ inside plasma has one
$x$--component and changes only lengthwise the axis $x$:
$$
\mathbf{E}=\{E_x(x,t),0,0\}.
$$

Under this configuration the electric field is perpendicular
to the boun\-da\-ry of plasma, which is situated in the plane $x=0$.

We will linearize the local equilibrium Fermi --- Dirac 
distribution $f_{eq}$
in regard to the undisturbed distribution $f_0(\E)$:
$$
f_{eq}=f_{0}(\E)+[\E_{F}(x,t)-\E]\delta(\E_{F}-\E),
$$
where $\delta(x)$ is the delta -- function of Dirac.

We also linearize the electron distribution function $f$ in terms of
absolute Fermi --- Dirac distribution
$f_{0}(\E)$:
$$
f=f_0(\E)+f_1(x,\mathbf{v},t).
\eqno{(1.4)}
$$

After the linearization of the equations (1.1)--(1.3) with the help
of (1.4) we obtain the following system of equations:

$$
\dfrac{\partial f_1}{\partial t}+v_{x}\dfrac{\partial f_1}{\partial
x}+\nu f_1(x, \mathbf{v}, t)=
$$

$$
=\delta(\E_{F}-\E) \big[e E_{x}(x,t)v_{x}+\nu[\E_{F}(x,t)-\E_{F}]\big] ,
\eqno{(1.5)}
$$
$$
\dfrac{\partial E_{x}(x,t)}{\partial x}=\dfrac{8\pi
e}{(2\pi\hbar)^{3}}\int f_1(x,\mathbf{v'}, t)d^3p' \eqno {(1.6)}
$$

From the law of preservation of number of particles
$$
\int f_{eq}d\Omega_F=\int f d\Omega_F
$$
we find:
$$
[\E_{F}(x,t)-\E_{F}]\int\delta(\E_{F}-\E)d^3p=\int f_1 d^3p.
\eqno{(1.7)}
$$

From the equation (1.5) it is seen that we should search for
the function $f_1$ in the form proportional to the delta -- function:
$$
f_1=\E_F \delta(\E_F-\E) H(x,\mu,t), \qquad \mu=\dfrac{v_x}{v}.
\eqno{(1.8)}
$$

The system of equations (1.5) and (1.6) with the help of (1.7) and (1.8)
can be transformed to the following form:
$$
\dfrac{\partial H}{\partial t}+v_{F}\mu \dfrac{\partial H}{\partial
x}+ \nu H(x,\mu,t)=
$$

$$
=\dfrac{ev_{F}\mu}{\E_{F}}E_{x}(x,t)+
\dfrac{\nu}{2}\int_{-1}^{1}H(x,\mu',t)d\mu',
$$

$$
\dfrac{\partial E_{x}(x,t)}{\partial x}=\dfrac{16\pi^{2}e\E_{F}m^2
v_{F}}{(2\pi \hbar)^3}\int_{-1}^{1}H(x,\mu',t)d\mu'.
$$

Further we introduce dimensionless function
$$
e(x,t)=\dfrac{ev_{F}}{\nu \E_{F}}E_{x}(x,t)
$$
and pass to dimensionless coordinate $x_1=x/l$ , where $\;l=v_F\tau$
is the mean free path of electrons, and we introduce dimensionless
time $t_1=\nu t$. We obtain the following system of equations:

$$
\dfrac{\partial H}{\partial t_1}+\mu \dfrac{\partial H}{\partial
x_1}+ \nu H(x_1,\mu,t_1)=
$$

$$
=\mu e(x_1,t_1)+\dfrac{1}{2}\int_{-1}^{1}H(x_1,\mu',t_1)d\mu',
\eqno{(1.9)}
$$

$$
\dfrac{\partial e(x_1,t_1)}{\partial
x_1}=\dfrac{3\omega_{p}^{2}}{2\nu^{2}}\int_{-1}^{1}H(x_1,\mu',t_1)d\mu'.
\eqno{(1.10)}
$$

Here $\omega_{p}$ is the electron (Langmuir) frequency 
of plasma oscillations,
$$
\omega_p^2=\dfrac{4\pi e^2N}{m},
$$
$N$ is the numerical density (concentration), $m$ is the electron mass.

We used the following well-known relation for degenerate plasma for
the derivation of the equations (1.9) and (1.10)
$$
\Big(\dfrac{v_F m}{\hbar}\Big)^3=3\pi^2 N.
$$

Let $k$ to be a dimensional wave number, and let us introduce
dimen\-si\-on\-less wave number
$k_1=k\dfrac{v_F}{\omega_p}$, then
$kx=\dfrac{k_1x_1}{\varepsilon}$, where
$\varepsilon=\dfrac{\nu}{\omega_p}$. We introduce the quantity
$\omega_1=\omega\tau=\dfrac{\omega}{\nu}$.

\begin{center}\bf
3.  BOUNDARY CONDITIONS STATEMENT
\end{center}

Let us outline the time variable of the functions $H(x_1,\mu,t_1)$
and $e(x_1,t_1)$, assuming
$$
H(x_1,\mu,t_1)=e^{-i\omega_1t_1}h(x_1,\mu),
\eqno{(2.1)}
$$
$$
e(x_1,t_1)=e^{-i\omega_1t_1}e(x_1).
\eqno{(2.2)}
$$

The system of equations (1.9) and (1.10) in this case will be
transformed to the following form:
$$
\mu\dfrac{\partial h}{\partial x_1}+(1-i\omega_1)h(x_1,\mu)= \mu
e(x_1)+\dfrac{1}{2}\int\limits_{-1}^{1}h(x_1,\mu')d\mu',
\eqno{(2.3)}
$$
$$
\dfrac{de(x_1)}{dx_1}=\dfrac{3\omega_p^2}{2\nu^2}
\int\limits_{-1}^{1}h(x_1,\mu')d\mu'.
\eqno{(2.4)}
$$

Further instead of $x_1,t_1$ we write $x,t$. We rewrite the system
of equations (2.3) and (2.4) in the form:
$$
\mu\dfrac{\partial h}{\partial x}+z_0h(x,\mu)= \mu
e(x)+\dfrac{1}{2}\int\limits_{-1}^{1}h(x,\mu')d\mu',
\eqno{(2.5)}
$$
$$
\dfrac{de(x)}{dx}=\dfrac{3}{2\varepsilon^2}
\int\limits_{-1}^{1}h(x,\mu')d\mu'. \eqno{(2.6)}
$$
Here
$$
z_0=1-i\omega_1=1-\frac{\omega}{\nu}=1-i\omega\tau.
$$

We consider the external electric field outside the plasma  is
perpendi\-cu\-lar to the plasma boundary and changes according to
the following law: $E_{0}\exp(-i\omega t)$.
This means that for the field inside plasma on the plasma
boundary the following condition is satisfied:
$$
e(0)=e_0.
\eqno{(2.7)}
$$

The non-flowing condition for the particle (electric current) flow
through the plasma
boundary means that
$$
\int\limits_{-1}^{1}\mu\,h(0,\mu)\,d\mu=0. \eqno{(2.8)}
$$

In the kinetic theory for the description of the surface properties
the accommodation coefficients are used often. Tangential momentum
and energy accommodation coefficients are the most--used. For the
problem considered the normal electron momentum accommodation under
the scat\-te\-ring on the surface has the most important significance.

The normal momentum accommodation coefficient
is defined by the follo\-wing relation:
$$
\alpha_p=\dfrac{P_i-P_r}{P_i-P_s}, \quad 0\leqslant \alpha_p
\leqslant 1, \eqno{(2.9)}
$$
where $P_i$ and $P_r$  are the flows of normal to the surface
momentum of incoming to the boundary and reflected from it electrons,
$$
P_i=\int\limits_{-1}^{0}\mu^2h(0,\mu)d\mu,
\eqno{(2.10)}
$$

$$
P_r=\int\limits_{0}^{1}\mu^2h(0,\mu)d\mu,
\eqno{(2.11)}
$$
quantity $P_s$ is the normal momentum flow for electrons
reflected from the surface which are in thermodynamic
equilibrium with the wall,
$$
P_s=\int\limits_{0}^{1}\mu^2h_s(\mu)d\mu,
\eqno{(2.12)}
$$
where the function
$$
h_s(\mu)=A_s,\quad 0<\mu<1,
$$
is the equilibrium distribution
function of the correspon\-ding electrons.
This function is to satisfy the condition
similar to the non-flowing condition:
$$
\int\limits_{-1}^{0}\mu h(0,\mu)d\mu+\int\limits_{0}^{1}\mu
h_s(\mu)d\mu=0.
\eqno{(2.13)}
$$

We are going to consider the relation between the normal momentum
accom\-mo\-da\-tion coefficient $\alpha_p$ and the diffuseness coefficient
$q$ for the case of specu\-lar and diffuse boundary conditions which
are written in the follo\-wing form:
\[
h(0,\mu)=(1-q) h(0,-\mu)+a_s, \quad 0<\mu<1. 
\eqno{(2.14)}
\]

Here $q$ is the diffuseness coefficient ($0\leqslant q\leqslant 1$),
$a_s$ is the quantity determined from the non-flowing condition.

From the non-flowing condition we derive
\[
\int\limits_{-1}^{1}\mu h(0,\mu)d\mu=\int\limits_{-1}^{0}\mu
h(0,\mu)d\mu+\int\limits_{0}^{1}\mu h(0,\mu)d\mu=0.
\]

In the second integral we replace the integrand according to the
right-hand side of the specular--diffuse boundary condition (2.14).
After that, using the obvious change of integration variable, we
obtain that
\[
a_s=-2q\int\limits_{-1}^{0}\mu h(0,\mu)d\mu.
\]

Let us use the boundary condition (2.13). Using the analogous to the
preceded line of reasoning we get:
\[
A_s=-2\int\limits_{-1}^{0}\mu h(0,\mu)d\mu.
\]
From the two last equations we find that
\[
a_s=qA_s. \eqno{(2.15)}
\]

Further we find the difference between two flows
\[
P_i-P_r=\int\limits_{-1}^{0}\mu^2
h(0,\mu)d\mu-\int\limits_{0}^{1}\mu^2 h(0,\mu)d\mu.
\]
In the second integral we use the boundary condition (2.14) again.
With the help of (2.15) we obtain that
\[
P_i-P_r=q\int\limits_{-1}^{0}\mu^2 h(0,\mu)d\mu-\int\limits_{0}^{1}\mu^2
a_sd\mu=
\]
\[
=q\int\limits_{-1}^{0}\mu^2 h(0,\mu)d\mu-q\int\limits_{0}^{1}\mu^2
A_sd\mu=qP_i-qP_s.
\]
Substituting the expressions obtained to the definition of the normal
mo\-men\-tum accommodation coefficient, we have:
$$
\alpha_p=\frac{P_i-P_r}{P_i-P_s}=\frac{qP_i-qP_s}{P_i-P_s}=q.
$$.
Thus, for specular -- diffuse boundary conditions normal momentum
accom\-mo\-da\-tion coefficient $\alpha_p$
coincides with the diffuseness coefficient $q$.

Equally with the specular -- diffuse boundary conditions another
vari\-ants of boundary conditions are used in kinetic theory as well.

In particular, accommodation boundary conditions are used widely.
They are divided into two types: diffuse -- accommodative and
specular -- accommodative boundary conditions (see \cite{14}).

We consider specular -- accommodative boundary conditions. For
the function $h$ this conditions will be written in the following form:
$$
h(0,\mu)=h(0,-\mu)+A_0+A_1\mu, \quad 0<\mu<1.
\eqno{(2.16)}
$$

If in (2.16) we assume $A_0=A_1=0$, then specular -- accommodative
boundary conditions pass into pure specular boundary conditions.

Coefficients $A_1$ and $A_2$ can be derived from the non-flowing
condition and the definition of the normal electron momentum
accommodation coef\-fi\-ci\-ent.

The problem statement is completed. Now the problem consists in
finding of such solution of the system of equations (2.5) and (2.6),
which satisfies the boundary conditions  (2.7)--(2.13). Further, with
the use of the solution of the problem, it is required to built the
profiles of the distribution function of the electrons moving to the
plasma surface, and profile of the electric field.

\begin{center}\bf
4. THE RELATION BETWEEN FLOWS
AND BOUNDARY CONDITIONS
\end{center}

First of all let us find expression which relates the constants
$A_0,\;A_1$ from the boundary condition (2.16). To carry this out
we will use the condition of non-flowing (2.12) of the particle flow
through the plasma boundary, which
we will write as a sum of two flows:
$$
N_0\equiv \int\limits_{0}^{1}\mu h(0,\mu)d\mu+\int\limits_{-1}^{0}
\mu h(0,\mu)d\mu=0.
$$

After evident substitution of the variable in the second
integral we obtain:
$$
N_0\equiv \int\limits_{0}^{1} \mu
\Big[h(0,\mu)-h(0,-\mu)\Big]d\mu=0.
$$

Taking into account the relation  (2.16), we obtain that
$$
A_0=-\frac{2}{3}A_1.
$$
With the help of this relation we can rewrite the condition (2.16)
in the following form:
$$
h(0,\mu)=h(0,-\mu)+A_1(\mu-\dfrac{2}{3}), \qquad 0<\mu<1.
\eqno{(3.1)}
$$

We consider the momentum flow of the electrons which are moving to
the boundary. According to (3.1) we have:
$$
P_i=P_r-\dfrac{1}{36}A_1.
\eqno{(3.2)}
$$

It is easy to see further that
$$
P_s=\frac{A_s}{3}. \eqno{(3.3)}
$$
With the help of the formulas (3.2) and (3.3) we will rewrite the
definition of
the accommodation coefficient (2.9) in the form:
$$
\alpha_pP_r-\alpha_p\dfrac{A_s}{3}+\dfrac{A_1}{36}(1-\alpha_p)=0.
\eqno{(3.4)}
$$

Let us consider the condition (2.13). We rewrite it in the following form:
$$
\dfrac{A_s}{2}+\int\limits_{-1}^{0}\mu h(0,\mu)d\mu=0.
$$

From this condition we obtain
$$
A_s=-2\int\limits_{-1}^{0}\mu h(0,\mu)d\mu=2\int\limits_{0}^{1}
\mu h(0,-\mu)d\mu.
$$

Using the condition (3.1), we then get
$$
A_s=2\int\limits_{0}^{1}\mu h(0,\mu)d\mu. \eqno{(3.5)}
$$

Now with the help of the second equality from (2.16) and (3.5) we
rewrite the relation (3.4) in the integral form:
$$
\alpha_p\int\limits_{0}^{1}\Big(\mu^2-\dfrac{2}{3}\mu\Big)h(0,\mu)d\mu=
-\dfrac{1}{36}(1-\alpha_p)A_1. \eqno{(3.6)}
$$
Now the boundary problem consists of the equations  (2.5) and (2.6)
and boundary conditions (2.7), (3.1) and (3.6).

\begin{center}\bf
5. SEPARATION OF VARIABLES AND CHARACTERISTIC SYSTEM
\end{center}

Application of the general Fourier method of the separation of
variables in several steps results in the following 
substitution \cite{Case}:
$$
h_\eta(x,\mu)=\exp(-\dfrac{z_0x}{\eta})\Phi(\eta,\mu),
\eqno{(4.1)}
$$

$$
e_\eta(x)=\exp(-\dfrac{z_0x}{\eta})E(\eta),
\eqno{(4.2)}
$$
where $\eta$ is the spectrum parameter or the parameter of
separation, which is complex in general.

We substitute the equalities (4.1) and (4.2) into 
the equations (2.5) and (2.6).
We obtain the following characteristic system of equations:
$$
z_0(\eta-\mu)\Phi(\eta,\mu)=\eta\mu E(\eta)+\dfrac{\eta}{2}
\int\limits_{-1}^{1}\Phi(\eta,\mu')d\mu',
\eqno{(4.3)}
$$
$$
E(\eta)=-\dfrac{3}{\varepsilon^2z_0}\cdot
\dfrac{\eta}{2}\int\limits_{-1}^{1}
\Phi(\eta,\mu')d\mu'.
\eqno{(4.4)}
$$

Let us introduce the designations:
$$
\gamma=\dfrac{\omega}{\omega_p}-1, \qquad \gamma\geq -1, \qquad
\eta_1^2=\dfrac{\varepsilon^2z_0}{3},\qquad
$$
$$
c=z_0\eta_1^2=\dfrac{\varepsilon^2z_0^2}{3}=\dfrac{1}{3}[\varepsilon -
i(1+\varepsilon)]^2,
$$
$$
\qquad z_0=1-i\dfrac{1+\gamma}{\varepsilon}.
$$

Substituting the integral from the equation (4.4) into (4.3), we
come to the following system of equations:
$$
(\eta-\mu)\Phi(\eta,\mu)=\dfrac{E(\eta)}{z_0}(\eta\mu-\eta_1^2),
\eqno{(4.5)}
$$
$$
-\eta_1^2E(\eta)=\dfrac{\eta}{2}\int\limits_{-1}^{1}
\Phi(\eta,\mu')d\mu'. \eqno{(4.6)}
$$

Here
$$
\eta_1^2=\dfrac{\varepsilon^2}{3}-i\dfrac{\varepsilon(1+\gamma)}{3}.
$$

Solution of the system (4.5) and (4.6) depends essentially on the
condi\-tion whether
the spectrum parameter $\eta$ belongs to the interval $-1<\eta<1$.
In connection with this the interval $-1<\eta<1$ we will call as
continuous spectrum of the characteristic system.

Let the parameter $\eta\in (-1,1)$. Then from the equations 
(4.5) and (4.6) in
the class of general functions we will find eigenfunction
corresponding to the continuous spectrum:
$$
\Phi(\eta,\mu)=F(\eta,\mu)\dfrac{E(\eta)}{z_0}, \eqno{(4.7)}
$$
where
$$
F(\eta,\mu)=P\dfrac{\mu\eta-\eta_1^2}{\eta-\mu}-c
\dfrac{\lambda(\eta)}{\eta}\delta(\eta-\mu). \eqno{(4.8)}
$$

In the equation (4.8) $\delta(x)$ is the delta--function of Dirac,
the symbol $Px^{-1}$ means the principal value of the integral under
integrating of the expression $x^{-1}$, the function $\lambda(z)$ is
called as dispersion function of the problem,
$$
\lambda(z)=1+\dfrac{z}{c}\int\limits_{-1}^{1}
\dfrac{\eta_1^2-z\mu}{\mu-z}d\mu.
\eqno{(4.9)}
$$

Functions (4.8) are  called eigenfunction of the continuous
spectrum, since the spectrum parameter $\eta$ fills out the
continuum $(-1,+1)$ compactly. The eigensolutions of the given
problem can be found from the equalities (4.7).

The dispersion
function $\lambda(z)$ we express in the terms of
the Case dispersion function \cite{Case}:
$$
\lambda(z)=1-\dfrac{1}{z_0}+\dfrac{1}{z_0}\Big(1-
\dfrac{z^2}{\eta_1^2}\Big)\lambda_c(z),
$$
where
$$
\lambda_c(z)=1+\dfrac{z}{2}\int\limits_{-1}^{1}\dfrac{d\tau}{\tau-z}=
\dfrac{1}{2}\int\limits_{-1}^{1}\dfrac{\tau\,d\tau}{\tau-z}
$$
is the Case dispersion function  \cite{Case}.

The boundary values of the dispersion function from above and below
the cut (interval $(-1,1)$) we define in the following way:
$$
\lambda^{\pm}(\mu)=\lim\limits_{\varepsilon\to 0, \varepsilon>0}
\lambda(\mu\pm i \varepsilon), \qquad \mu\in (-1,1).
$$

The boundary values of the dispersion function from above and below
the cut are calculated according to the Sokhotzky formulas
$$
\lambda^{\pm}(\mu)=\lambda(\mu)\pm \dfrac{i \pi\mu}
{2\eta_1^2z_0}(\eta_1^2-\mu^2),\quad -1<\mu<1,
$$
from where
$$
\lambda^+(\mu)-\lambda^-(\mu)=\dfrac{i \pi}{\eta_1^2z_0}
\,\mu(\eta_1^2-\mu^2),
$$
$$
\dfrac{\lambda^+(\mu)+\lambda^-(\mu)}{2}=\lambda(\mu),\quad-1<\mu<1,
$$
where
$$
\lambda(\mu)=1+\dfrac{\mu}{2\eta_1^2z_0} \int\limits_{-1}^{1}
\dfrac{\eta_1^2-\eta^2}{\eta-\mu}\,d\eta,
$$
and the integral in this equality is understood as singular in terms
of the principal value by Cauchy. Besides that, the function
$\lambda(\mu)$ can be represented in the following form:
$$
\lambda(\mu)=1-\dfrac{1}{z_0}+
\dfrac{1}{z_0}\Big(1-\dfrac{\mu^2}{\eta_1^2}\Big)\lambda_c(\mu),
$$
$$
 \lambda_c(\mu)=1+\dfrac{\mu}{2}\ln\dfrac{1-\mu}{1+\mu}.
$$

\begin{center}\bf
6. EIGENFUNCTIONS OF THE DISCRETE SPECTRUM
\end{center}

According to the definition, the discrete spectrum of the
characteristic equation is a set of zeroes of the dispersion
equation
$$
\dfrac{\lambda(z)}{z}=0. \eqno{(5.1)}
$$

We start to search zeroes of the equation (5.1). Let us take Laurent
series of the dispersion function:
$$
\lambda(z)=\lambda_\infty+\dfrac{\lambda_2}{z^2}+
\dfrac{\lambda_4}{z^4}+\cdots,\qquad |z|>1.
\eqno{(5.2)}
$$

Here
$$
\lambda_\infty \equiv\lambda(\infty)=
1-\dfrac{1}{z_0}+\dfrac{1}{3z_0\eta_1^2},
$$
$$
\lambda_2=-\dfrac{1}{z_0}\Big(\dfrac{1}{3}-\dfrac{1}{5\eta_1^2}\Big),
$$
$$
\lambda_4=-\dfrac{1}{z_0}\Big(\dfrac{1}{5}-\dfrac{1}{7\eta_1^2}\Big).
$$

We express these parameters through the parameters $\gamma$ and
$\varepsilon$:
$$
\lambda_\infty \equiv\lambda(\infty)=
\dfrac{2\gamma+i\varepsilon+\gamma(\gamma+i\varepsilon)}{(1+\gamma+
i\varepsilon)^2},
$$
$$
\lambda_2=-\dfrac{9+5i\varepsilon(1+\gamma+i\varepsilon)}{15(1+\gamma+
i\varepsilon)^2},
$$
$$
\lambda_4=-\dfrac{15+7i\varepsilon(1+\gamma+i\varepsilon)}{35(1+\gamma+
i\varepsilon)^2}.
$$

It is easy seen that the dispersion function (4.9) in collisional plasma
(i.e. when $\varepsilon>0$) in the infinity has the value which doesn't
equal to zero:
$\lambda_\infty=\lambda(\infty)\ne 0$.

Hence, the dispersion equation has infinity as a zero $\eta_i=\infty$,
to which the discrete eigensolutions of the given system correspond:
$$
h_\infty(x,\mu)=\dfrac{\mu}{z_0},\qquad\;e_\infty(x)=1.
$$

This solution is naturally called as mode of Drude. It describes the
volume conductivity of metal, considered by Drude
(see, for example, \cite{16}).

Let us consider the question of the plasma mode existence in
details. We find finite complex zeroes of the dispersion function.
We use the principle of argument. We take the contour (see Fig. 1)
$$
\Gamma_\varepsilon^+=\Gamma_R\cup \gamma_\varepsilon,
$$
which is passed in the positive direction and which bounds the
biconnected domain $D_R$. This contour consists of the circumference
$$
\Gamma_R=\{z:\quad |z|=R,\quad\;R=\dfrac{1}{\varepsilon},\quad
\varepsilon>0\},
$$
and the
contour $\gamma_\varepsilon$, which includes the cut $[-1,+1]$, and
stands at the distance of $\varepsilon$ from it.

Let us note that the dispersion function has not any poles in the
domain $D_R$.
Then according to the principle of argument the number \cite{17} of
zeroes $N$ in the domain $D_R$ equals to:
$$
N=\dfrac{1}{2\pi
i}\oint\limits_{\Gamma_\varepsilon}d\,\ln\lambda(z).
$$

\begin{figure}
\begin{center}
\includegraphics{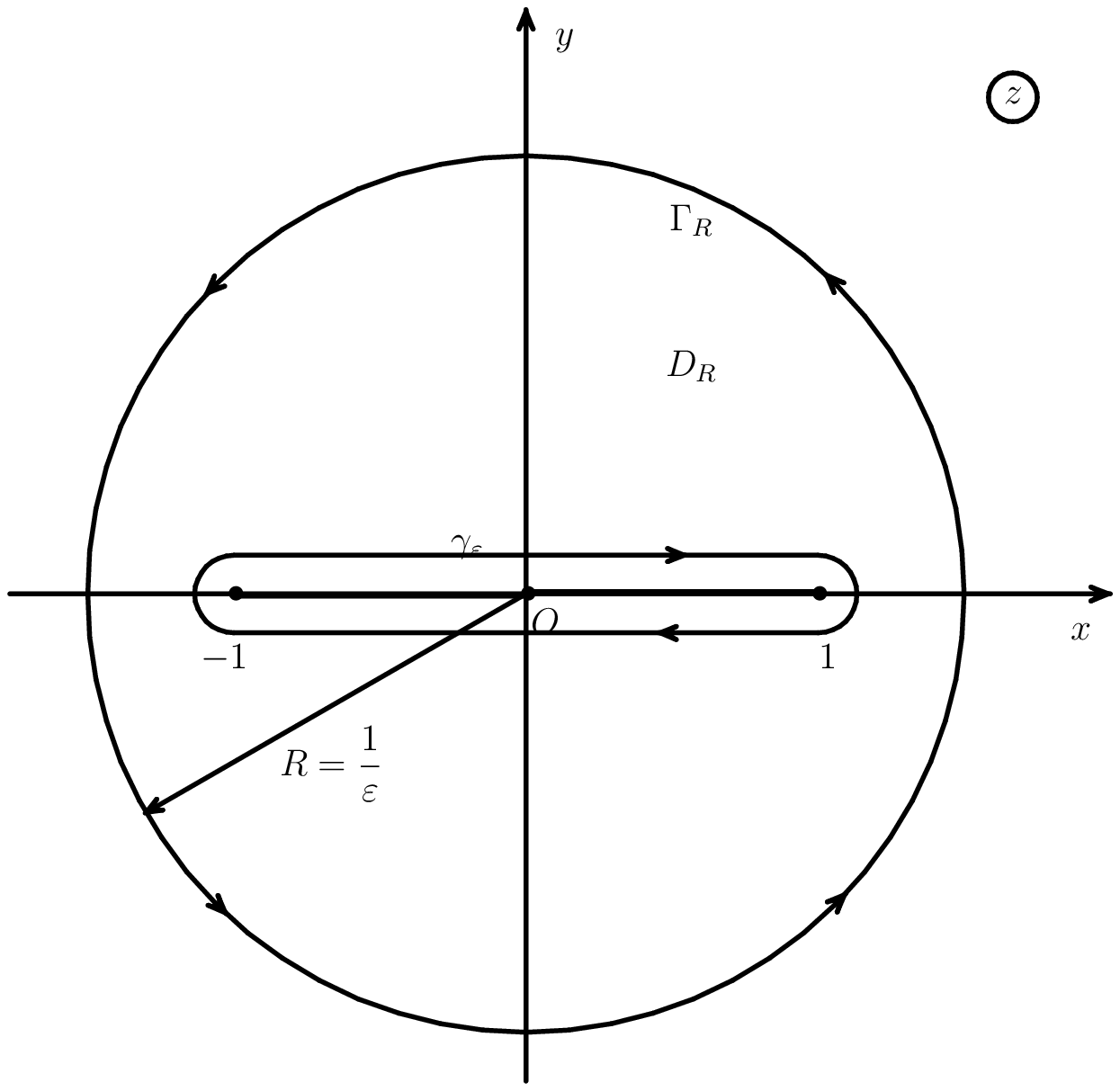}
\begin{center}
{\bf Fig. 1.}
\end{center}
\end{center}
\end{figure}

Considering the limit in this equality when $\varepsilon\to 0$ and
taking into account that the dispersion function is analytic in the
neighbourhood of the infinity, we obtain that
$$
N=\dfrac{1}{2\pi i}\int\limits_{-1}^{1}d\,\ln
\lambda^+(\tau)-\dfrac{1}{2\pi i}\int\limits_{-1}^{1}d\,\ln
\lambda^-(\tau)
$$
$$=\dfrac{1}{2\pi i}\int\limits_{-1}^{1}d\,\ln
\dfrac{\lambda^+(\mu)}{\lambda^-(\mu)}.
$$

So, we obtained that
$$
N=\dfrac{1}{2\pi i}\int\limits_{-1}^{1}d\,\ln
\dfrac{\lambda^+(\tau)}{\lambda^-(\tau)}.
$$
We divide this integral into two integrals by segments $[-1,0]$
and $[0,1]$. In the first integral by the segment $[-1,0]$ we
carry out replacement of variable
$ \tau \rightarrow -\tau $. Taking into account that
$\lambda^+(-\tau)=\lambda^-(\tau)$, we obtain that
$$
N=\dfrac{1}{2\pi i}\int\limits_{-1}^{1}d\,\ln
\dfrac{\lambda^+(\tau)}{\lambda^-(\tau)}= \dfrac{1}{\pi
i}\int\limits_{0}^{1}d\,\ln
\dfrac{\lambda^+(\tau)}{\lambda^-(\tau)}=\dfrac{1}{\pi}\arg
\dfrac{\lambda^+(\tau)}{\lambda^-(\tau)}\Bigg|_0^1.
\eqno{(5.3)}
$$

Here under symbol $\arg G(\tau)=\arg \dfrac{\lambda^+(\tau)}
{\lambda^-(\tau)}$ we understand the regular branch of the argument,
fixed in zero with the condition: $\arg G(0)=0$.

We consider the curve
$$
\gamma=\{z:\;z=G(\tau),\quad\;0\leqslant \tau\leqslant+1\},
$$
where
$$
G(\tau)=\dfrac{\lambda^+(\tau)}{\lambda^-(\tau)}.
$$

It is obvious that
$$
G(0)=1,\qquad\; \lim\limits_{\tau\to +1}G(\tau)=1.
$$

Consequently, according to (5.3), the number of values $N$ equals
to doubled number of turns of the curve $\gamma$ around the point
of origin, i.e.
$$
N=2\varkappa(G),
$$
where
$$
\varkappa(G)={\rm Ind}_{[0,+1]}G(\tau)
$$
is the index of the function $G(\tau)$.

Thus, the number of zeroes of the dispersion function, which are
situated in complex plane outside of the segment $[-1,1]$ of the
real axis, equals to doubled index of the function $G(\tau)$,
calculated on the "semi-segment"\, $[0,+1]$.

Let us single real and imaginary parts of the function $G(\mu)$ out.
At first, we represent the function $G(\mu)$ in the form:
$$
G(\mu)=\dfrac{(z_0-1)\eta_1^2+(\eta_1^2-\mu^2)\lambda_0(\mu)+
is(\mu)(\eta_1^2-\mu^2)}{(z_0-1)\eta_1^2+
(\eta_1^2-\mu^2)\lambda_0(\mu)- is(\mu)(\eta_1^2-\mu^2)}.
$$
where
$$
s(\mu)=\dfrac{\pi}{2}\mu, \qquad
$$
$$
\lambda(\mu)=1-\dfrac{1}{z_0}+
\dfrac{1}{z_0}\Big(1-\dfrac{\mu^2}{\eta_1^2}\Big)\lambda_c(\mu),
$$
and
$$
\lambda_c(\mu)=1+\dfrac{\mu}{2}\ln\dfrac{1-\mu}{1+\mu}
$$
is the dispersion function of Case, calculated on the cut
(i.e., in the interval $(-1,1)$).

Taking into account that
$$
z_0-1=-i\dfrac{\omega}{\nu}=-i\dfrac{1+\gamma}{\varepsilon}, \qquad
\eta_1^2=\dfrac{\varepsilon c}{3}=\dfrac{\varepsilon^2}{3}-
i\dfrac{\varepsilon(1+\gamma)}{3},
$$
$$
(z_0-1)\eta_1^2=-\dfrac{(1+\gamma)^2}{3}-i\dfrac{\varepsilon(1+\gamma)}
{3},
$$
we obtain
$$
G(\mu)=\dfrac{P^-(\mu)+iQ^-(\mu)}{P^+(\mu)+iQ^+(\mu)},
$$
where
$$
P^{\pm}(\mu)=(1+\gamma)^2-\lambda_0(\mu)(\varepsilon^2-3\mu^2)\pm
\varepsilon(1+\gamma)s(\mu),
$$
$$
Q^{\pm}(\mu)=\varepsilon(1+\gamma)(1+\lambda_0(\mu))\pm
s(\mu)(\varepsilon^2-3\mu^2).
$$

Now we can easily single real and imaginary parts of the function
$G(\mu)$ out:
$$
G(\mu)=\dfrac{g_1(\mu)}{g(\mu)}+i\dfrac{g_2(\mu)}{g(\mu)}.
$$

Here
$$
g(\mu)=[P^+(\mu)]^2+[Q^+(\mu)]^2=[(1+\gamma)^2+\lambda_0(3\mu^2-
\varepsilon^2)-$$$$-\varepsilon(1+\gamma)s]^2+
[\varepsilon(1+\gamma)(1+\lambda_0)- s(3\mu^2-\varepsilon^2)]^2,
$$

$$
g_1(\mu)=P^+(\mu)P^-(\mu)+Q^+(\mu)Q^-(\mu)=[(1+\gamma)^2+
\lambda_0(3\mu^2-$$$$\qquad\qquad-\varepsilon^2)]^2-
\varepsilon^2(1+\gamma)^2[s^2-(1+\lambda_0)^2]-
(3\mu^2-\varepsilon^2)^2s^2,
$$

$$
g_2(\mu)=P^+(\mu)Q^-(\mu)-P^-(\mu)Q^+(\mu)=
2s[(1+\gamma)^2(3\mu^2-\varepsilon^2)+$$$$
\qquad\qquad+\lambda_0(3\mu^2-
\varepsilon^2)^2+\varepsilon^2(1+\gamma)^2(1+\lambda_0)],
$$

We consider (see Fig. 2) the curve $L$, which is defined in
implicit form by the following parametric equations:
$$
L=\{(\gamma,\varepsilon): \qquad g_1(\mu;\gamma,\varepsilon)=0,\;\quad
g_2(\mu;\gamma,\varepsilon)=0, \quad 0\le \mu\le 1\},
$$
and which lays in the plane of the parameters of the problem
$(\gamma,\varepsilon)$,
and when passing through this curve the index of the function $G(\mu)$
at the positive "semi-segment"\, $[0,1]$ changes stepwise.

From the equation $g_2=0$ we find:
$$
(1+\gamma)^2=-\dfrac{\lambda_0(\mu)(3\mu^2-\varepsilon^2)}
{3\mu^2+\varepsilon^2\lambda_0(\mu)}.
\eqno{(5.4)}
$$

Now from the equation $g_1=0$ with the help of (5.4) we find that
$$
\varepsilon=\sqrt{L_2(\mu)}, \eqno{(5.5)}
$$
where
$$
L_2(\mu)=-\dfrac{3\mu^2s^2(\mu)}{\lambda_0(\mu)
[s^2(\mu)+(1+\lambda_0(\mu))^2]}.
$$

Substituting (5.5) into (5.4), we obtain:
$$
\gamma=-1+\sqrt{L_1(\mu)}, \eqno{(5.6)}
$$
where
$$
L_1(\mu)=-\dfrac{3\mu^2[s^2(\mu)+
\lambda_0(\mu)(1+\lambda_0(\mu))]^2}{\lambda_0(\mu)[s^2(\mu)+
(1+\lambda_0(\mu))^2]}.
$$

Functions (5.5) and (5.6) determine the curve $L$ which is the
border if the domain $D^+$ (we designate the external area to the
domain as $D^-$) in explicit parametrical form (see Fig. 2). As in
the work \cite{18} we can prove that if $(\gamma,\varepsilon)\in
D^+$, then $\varkappa(G)={\rm Ind}_{[0,+1]} G(\mu)=1$ (the curve $L$
encircles the point of origin once), and if $(\gamma,\varepsilon)\in
D^-$, then $\varkappa(G)={\rm Ind}_{[0,+1]} G(\mu)=0$ (the curve $L$
doesn't encircle the point of origin).

We note, that in the work \cite{18} 
the method of analysis of boundary regime when
$(\gamma,\varepsilon)\in L$ was developed.

From the expression (3.2) one can see that the number of zeroes of
the dispersion function equals to two if $(\gamma,\varepsilon)\in
D^+$, and equals to zero if $(\gamma,\varepsilon)\in D^-$.

\begin{figure}
\begin{center}
\includegraphics[width=16cm, height=10cm]{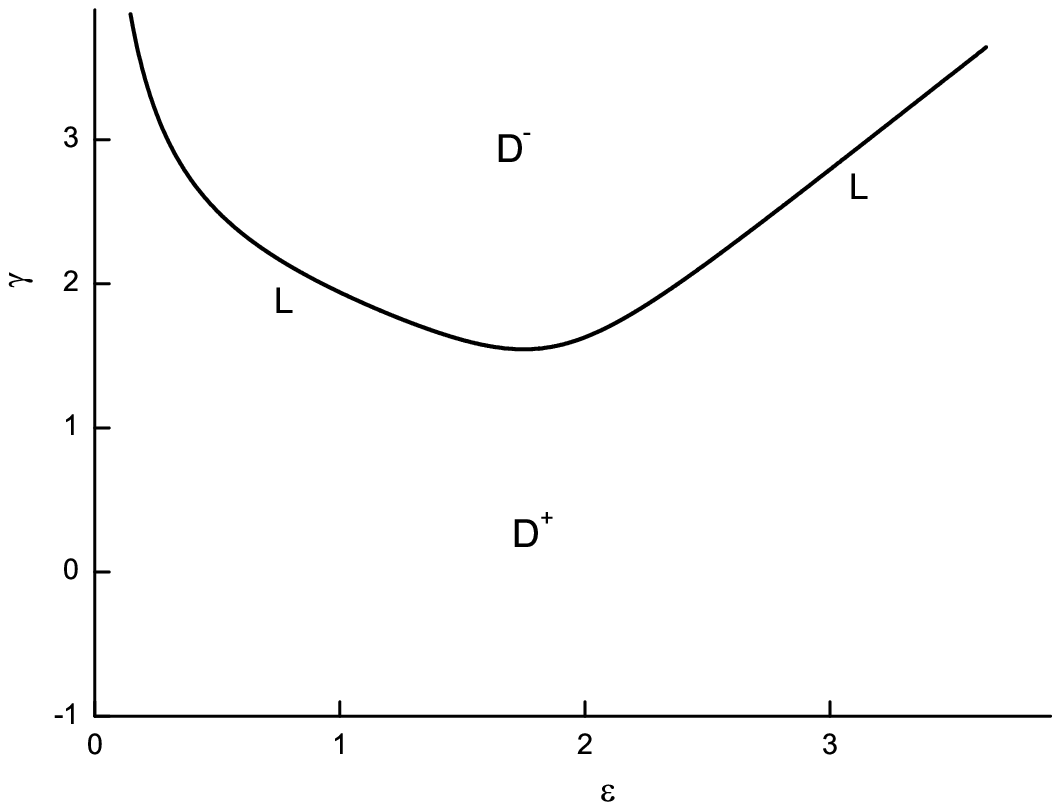}
\end{center}
\begin{center}
{\bf Fig. 2.}
\end{center}
\end{figure}

Since the dispersion function is even its zeroes differ from each
other by sign. We designate these zeroes as following $\pm \eta_0$,
by $\eta_0$ we take the zero which satisfies the condition $\Re
\eta_0>0$. The following solution corresponds to the zero $\eta_0$
$$
h_{\eta_0}(x,\mu)=\exp\Big(-\dfrac{z_0x}{\eta_0}\Big)
\dfrac{E_2}{z_0}\dfrac{\eta_0\mu-\eta_1^2}{\eta_0-\mu},
\eqno{(5.7)}
$$

$$
e_{\eta_0}(x)=\exp\Big(-\dfrac{z_0x}{\eta_0}\Big)E_2.
\eqno{(5.8)}
$$

From the equalities (5.7) and (5.8) it follows that the following
solution corresponds to zero $-\eta_0$:
$$
h_{-\eta_0}(x,\mu)=\exp(-\dfrac{z_0x}{\eta_0})
\dfrac{E_1}{z_0}\dfrac{\eta_0\mu+\eta_1^2}{\eta_0+\mu},
\eqno{(5.9)}
$$

$$
e_{-\eta_0}(x)=\exp(\dfrac{z_0x}{\eta_0})E_1.
\eqno{(5.10)}
$$

This solution is naturally called as mode of Debay (this is plasma
mode). In the case of low frequencies it describes well-known
screening of Debay \cite{Abrikosov}. The external field penetrates
into plasma on the depth of $r_D,\; r_D$ is the raduis of Debay.
When the external field frequencies are close to Langmuir
frequencies, the mode of Debay describes plasma oscillations (see,
for instance, \cite{Abrikosov,16}).

\begin{center}\bf
 7. EXPANSION IN THE TERMS OF EIGEN FUNCTIONS
\end{center}

We will seek for the solution of the system of equations (2.5)
and (2.6) with boundary conditions (3.1), (3.6) and (2.7) in the form
of linear combi\-na\-tion of discrete eigen solutions of the
characteristic system and integral taken over continuous spectrum of
the system. Let us prove that the following theorem is true.

\textbf{Theorem 7.1}. \emph{System of equations (2.5) and (2.6) with
boundary conditions (3.1), (3.6) and (2.7) has a unique solution,
which can be presented as an expansion by eigen functions of the
characteristic system:}
$$
h(x,\mu)=\dfrac{E_\infty}{z_0}\mu+\dfrac{E_0}{z_0}
 \dfrac{\eta_0\mu-\eta_1^2}{\eta_0-\mu}
 \exp\Big(-\dfrac{z_0x}{\eta_0}\Big)+
 $$
 $$
 +\dfrac{1}{z_0}
 \intl_{0}^{1}\exp\Big(-z_0\dfrac{x}{\eta}\Big)
 F(\eta,\mu)E(\eta)\,d\eta,
 \eqno{(6.1)}
 $$
$$
 e(x)=E_\infty+E_0\exp\Big(-z_0\dfrac{x}{\eta}\Big)+
 \intl_{0}^{1}
 \exp\Big(-z_0\dfrac{x}{\eta}\Big)E(\eta)\,d\eta.
 \eqno{(6.2)}
 $$

\emph{Here} $E_0$ \emph{and} $E_\infty$ \emph{is unknown
coefficients corresponding to the disc\-re\-te spectrum} ($E_0$ \emph{is
the amplitude of Debay}, $E_1$ \emph{is the amplitude of Drude),
}$E(\eta)$ \emph{is unknown function, which is called as coefficient
of discrete spect\-rum}.

When $(\gamma,\varepsilon)\in D^-$ in expansions (6.1) and (6.2) we
should take $E_0=0$. Then from (6.1) and (6.2) we obtain:
$$
h(x,\mu)=\dfrac{E_\infty}{z_0}\mu+\dfrac{1}{z_0}
 \intl_{0}^{1}\exp\Big(-z_0\dfrac{x}{\eta}\Big)
 F(\eta,\mu)E(\eta)\,d\eta,
\eqno{(6.1')}
$$
$$
 e(x)=E_\infty+ \intl_{0}^{1}
\exp\Big(-z_0\dfrac{x}{\eta}\Big)E(\eta)\,d\eta.
 \eqno{(6.2')}
$$

Further we will consider the following case $(\gamma,\varepsilon)\in
D^+$.

Our purpose is to find the coefficient of the continuous spectrum,
coeffi\-ci\-ents of the discrete spectrum and to built expressions for
electron distri\-bu\-tion function at the plasma surface and electric
field.

\textbf{Proof.} We substitute the expression (6.1) into the
boundary condition (3.1). We obtain the following equation in the
interval $0<\mu<1$:
 $$
2\dfrac{E_\infty}{z_0}\mu+\dfrac{E_0}{z_0}\varphi_0(\mu)+$$$$+
\dfrac{1}{z_0}\intl_{0}^{1}\Big[F(\eta,\mu)-
 F(\eta,-\mu)\Big]E(\eta)d\eta=A_1\Big(\mu-\dfrac{2}{3}\Big),
 \eqno{(6.3)}
 $$
where the following designation is introduced
 $$
 \varphi_0(\mu)=\dfrac{\eta_1^2-\eta_0\mu}{\mu-\eta_0}+
 \dfrac{\eta_1^2+\eta_0\mu}{\mu+\eta_0}.
 $$

 Extending the function $E(\eta)$ into the interval $(-1,0)$ evenly,
 so that $E(\eta)=E(-\eta)$, and extending
 the equation into the interval $(-1;1)$ unevenly, we transform
 the equation (6.3) to the form:
 $$
2\dfrac{E_\infty}{z_0}\mu+\dfrac{E_0}{z_0}\varphi_0(\mu)+
\dfrac{1}{z_0}\intl_{-1}^{1}F(\eta,\mu)E(\eta)d\eta
$$
$$=
A_1\Big(\mu-\dfrac{2}{3}\sign\mu\Big).
 \eqno{(6.4)}
 $$

We substitute the eigen functions of the continuous spectrum into the
equation (6.4). We obtain the singular integral equation with the
Cauchy kernel in the interval $(-1,1)$:
 $$
2E_\infty\mu+E_0\varphi_0(\mu)+\intl_{-1}^{1}
 \dfrac{\eta\mu-\eta_1^2}{\eta-\mu}E(\eta)d\eta-
$$
$$
-2\eta^2_1z_0\dfrac{\lambda(\mu)}{\mu}E(\mu)-
 z_0A_1\mu=-\dfrac{2}{3}z_0A_1\sign \mu.
\eqno{(6.5)}
 $$

\begin{center}\bf
8. SOLUTION OF THE SINGULAR EQUATION
\end{center}

 We introduce the auxiliary function
 $$
 M(z)=\intl_{-1}^{1}\dfrac{\eta z-\eta_1^2}{\eta-z}
 E(\eta)d\eta,
\eqno{(7.1)}
 $$
 the boundary values of which on the real axis above and below it
 are related by the Sohotzky formulas:
 $$
 M^+(\mu)-M^-(\mu)=2\pi i (\mu^2-\eta_1^2)E(\mu).
\eqno{(7.2)}
 $$
$$
\dfrac{M^+(\mu)+M^-(\mu)}{2}=M(\mu),\qquad -1<\mu<+1,
\eqno{(7.3)}
$$
where
$$
M(\mu)=\intl_{-1}^{1}\dfrac{\eta \mu-\eta_1^2}{\eta-\mu}
E(\eta)d\eta,
$$
and the singular integral in this equality is understood as singular
in the sense of principal value of Cauchy.

With the help of the equalities (7.2) and (7.3) and the similar
equalities for the dispersion and auxiliary functions we reduce the
equation (6.5) to the boundary condition of the problem of
determination of analytic function by its jump on the contour:
 $$
 \lambda^+(\mu)[M^+(\mu)+2E_\infty\mu+E_0\varphi_0(\mu)-z_0A_1\mu]-
 $$$$-
\lambda^-(\mu)[M^-(\mu)+2E_\infty\mu+E_0\varphi_0(\mu)-z_0A_1\mu]=$$$$
=-\dfrac{2i\pi }{3\eta_1^2z_0}z_0A_1\mu(\eta_1^2-\mu^2)\sign \mu,
\quad -1<\mu<1.
 $$

This equation has general solution (see \cite{17}):
$$
\lambda(z)[M(z)+E_0\varphi(z)+2E_\infty z_0-z_0A_1z]
=$$$$=\dfrac{2}{3}\dfrac{z_0A_1}{2\eta_1^2z_0}\int\limits_{-1}^{1}
\dfrac{\mu(\mu^2-\eta_1^2)\sign \mu}{\mu-z}d\mu+C_1z,
$$
where $C_1$ is an arbitrary constant.

Let us introduce auxiliary function
$$
T(z)=\dfrac{1}{2\eta_1^2z_0}\int\limits_{-1}^{1}
\dfrac{\mu(\mu^2-\eta_1^2)\sign \mu}{\mu-z}d\mu.
$$

Then the general solution will be written in the form:
$$
\lambda(z)[M(z)+E_0\varphi_0(z)+2E_\infty
z_0-z_0A_1z]$$$$=\dfrac{2}{3}z_0A_1T(z)+C_1z,
$$

From the general solution we can easy find the function $M(z)$:
$$
M(z)=-2E_\infty z-E_0\varphi_0(z)+z_0A_1z
$$
$$+
\dfrac{2}{3}z_0A_1\dfrac{T(z)}{\lambda(z)}+\dfrac{C_1z}{\lambda(z)}.
\eqno{(7.4)}
$$

Let us eliminate the pole of the solution (7.4) in the infinity. We
get that
$$
C_1=(2E_\infty-z_0A_1)\lambda_\infty.
$$

Now we eliminate the poles of the solution (7.4) in the points $\pm
\eta_0$. We single out items which contain polar singularity in the
points $z=\pm \eta_0$ in the right-hand side of the solution (7.4):
$$
E_0\dfrac{\eta_0z-\eta_1^2}{\eta_0-z}=
\dfrac{C_1z+\frac{2}{3}z_0A_1T(z)}{\lambda(z)}
$$
and
$$
E_0\dfrac{\eta_0z+\eta_1^2}{\eta_0+z}=
\dfrac{C_1z+\frac{2}{3}z_0A_1T(z)}{\lambda(z)}.
$$
From the last two relations one can see that poles in the points
$z=\pm \eta_0$ can be eliminated by one equality since the functions
which compose the general solution are uneven:
$$
z_0A_1=\dfrac{E_0\lambda'(\eta_0)(\eta_1^2-\eta_0^2)-
2E_\infty\lambda_\infty\eta_0}
{(2/3)T(\eta_0)-\lambda_\infty \eta_0}.
\eqno{(7.5)}
$$

We find the coefficient of the continuous spectra from the Sohotzky
formula (7.2):
$$
E(\eta)=\dfrac{1}{2\pi i
(\mu^2-\eta_1^2)}\Big[M^+(\mu)-M^-(\mu)\Big].
\eqno{(7.6)}
$$

We substitute the expansion (6.1) for the function $h(x,\mu)$ into
the integral boundary condition (3.6). We derive the following
equation:
$$
\frac{1}{36}E_\infty+E_0
m(\eta_0)+\int\limits_{0}^{1}m(\eta)E(\eta)d\eta= $$$$
=-\dfrac{1}{36}z_0A_1\dfrac{1-\alpha_p}{\alpha_p}.
\eqno{(7.7)}
$$
We introduced the following designations in (7.7):
$$
m(\pm \eta_0)=\int\limits_{0}^{1}\Big(\mu^2-\dfrac{2}{3}\mu\Big)
F(\pm \eta_0,\mu)d\mu,\quad
$$
and
$$
m(\eta)=\int\limits_{0}^{1}\Big(\mu^2-\dfrac{2}{3}\mu\Big)
F(\eta,\mu)d\mu.
$$

The difference between boundary values of the auxiliary function
$M(z)$ from the expression (7.6) can be found with the help of the
general solution (7.4):
$$
E(\eta)=\dfrac{1}{2\pi i (\eta^2-\eta_1^2)}
\Bigg\{\dfrac{2}{3}z_0A_1\Big[\dfrac{T^+(\eta)}{\lambda^+(\eta)}-
\dfrac{T^-(\eta)}{\lambda^-(\eta)}\Big]+
$$
$$
+(2E_\infty-z_0A_1)\lambda_\infty \eta
\Big[\dfrac{1}{\lambda^+(\eta)}-
\dfrac{1}{\lambda^-(\eta)}\Big]\Bigg\}.
\eqno{(7.8)}
$$

Let us note, that when passing the positive part of the cut $(0,1)$
the functions $T(z)$ and $\lambda(z)$ make jumps, which differs from
each other only by sign. Indeed, let us represent the formula for
$T(z)$ in the following form:
$$
T(z)=\dfrac{1}{2\eta_1^2z_0}\int\limits_{0}^{1}\mu
(\mu^2-\eta_1^2)\Bigg[\dfrac{1}{\mu-z}-\dfrac{1}{\mu+z}\Bigg]d\mu=
$$
$$
=\dfrac{z}{\eta_1^2z_0}\int\limits_{0}^{1}\dfrac{\mu(\mu^2-\eta_1^2)}
{\mu^2-z^2}d\mu,
$$
or
$$
T(z)=\dfrac{z}{2\eta_1^2z_0}\int\limits_{0}^{1}
(\mu^2-\eta_1^2)\Bigg[\dfrac{1}{\mu-z}+\dfrac{1}{\mu+z}\Bigg]d\mu=
\dfrac{z}{\eta_1^2z_0}\int\limits_{0}^{1}\dfrac{\mu(\mu^2-\eta_1^2)}
{\mu^2-z^2}d\mu.
$$

This integral can be easily calculated in explicit form. On the cut
it is calculated by the following formula:
$$
T(\eta)=\dfrac{\eta}{2\eta_1^2z_0}\Big[1+(\eta^2-\eta_1^2)
\ln\Big(\dfrac{1}{\eta^2}-1\Big)\Big], \qquad -1<\eta<+1.
$$

Now from the Sokhotzky formula for difference of boundary values we
obtain that when $0<\eta<1$:
$$
\lambda^+(\eta)-\lambda^-(\eta)=\lambda(\eta)\pm \dfrac{i\pi
\eta(\eta_1^2-\eta^2)}{2\eta_1^2z_0},
$$
$$
T^+(\eta)-T^-(\eta)=T(\eta)\pm \dfrac{i\pi \eta
(\eta^2-\eta_1^2)}{2\eta_1^2z_0}.
$$

Now one can find that
$$
T^+(\eta)\lambda^-(\eta)-T^-(\eta)\lambda^+(\eta)=
2(T(\eta)+\lambda(\eta))\cdot \dfrac{i\pi \eta
(\eta^2-\eta_1^2)}{2\eta_1^2z_0},
$$
$$
\lambda^-(\eta)-\lambda^+(\eta)= 2 \dfrac{i\pi \eta
(\eta^2-\eta_1^2)}{2\eta_1^2z_0}.
$$

Let us introduce an integral:
$$
T_0(z)= \dfrac{1}{2\eta_1^2z_0}\int\limits_{0}^{1}
\dfrac{\eta^2-\eta_1^2}{\eta-z}d\eta.
$$

It is obvious, that in the comlex plane, this integral is calculated
by the formula:
$$
T_0(z)=\dfrac{z}{2\eta_1^2z_0}\Big[\dfrac{1}{2}+z+(z^2-\eta_1^2)
\ln\Big(\dfrac{1}{z}-1\Big)\Big].
$$

With the help of this function we present the dispersion function in
the following form:
$$
\lambda(z)=1-zT_0(z)+zT_0(-z),
$$
we also express the function $T(z)$ in terms of this integral:
$$
T(z)=zT_0(z)+zT_0(-z).
$$
Sum of the two last expressions equals to:
$$
\lambda(z)+T(z)=1+2zT_0(-z).
$$

We note, that the integral $T(-z)$ is not singular on the cut
$0<\eta<1$. The sum $\lambda(\eta)+T(\eta)$ on the cut $0<\eta<1$ is
calculated in explicit form without applying to integrals:
$$
\lambda(\eta)+T(\eta)=1+\dfrac{1}{2\eta_1^2z_0}\Big[\eta-
2\eta^2+2\eta(\eta^2-\eta_1^2)\ln\Big(\dfrac{1}{\eta}+1\Big)\Big].
$$

From the relations (7.6) and (7.8) we find the coefficient of
continuous spectra in explicit form:

$$
E(\eta)=\dfrac{1}{2\eta_1^2z_0\lambda^+(\eta)
\lambda^-(\eta)}\Big[2E_\infty\lambda_\infty\eta^2+
$$
$$
+z_0A_1\frac{2}{3}\eta(1+2zT_0(-z))-
\lambda_\infty\eta^2\Big],
$$
or
$$
E(\eta)=\dfrac{1}{2\eta_1^2z_0\lambda^+(\eta)\lambda^-(\eta)}
\Big[(2E_\infty-1)\lambda_\infty\eta^2+
$$
$$
+z_0A_1\Big(\frac{2}{3}\eta(1+2zT_0(-\eta)
)\Big)\Big].
\eqno{(7.9)}
$$

Let us calculate the integrals $m(\pm \eta_0)$ and $m(\eta)$ in
explicit form. The integrals $m(\pm \eta_0)$ can be calculated
easily:
$$
m(\eta_0)=(\eta_1^2-\eta_0^2)\Bigg[\Big(-\dfrac{1}{6}+\eta_0\Big)+
\Big(\eta_0^2-\dfrac{2}{3}\eta_0\Big)\ln
\Big(1-\dfrac{1}{\eta_0}\Big)\Bigg],
$$
$$
m(-\eta_0)=(\eta_1^2-\eta_0^2)\Bigg[\Big(-\dfrac{1}{6}-\eta_0\Big)+
\Big(\eta_0^2+\dfrac{2}{3}\eta_0\Big)\ln
\Big(1+\dfrac{1}{\eta_0}\Big)\Bigg].
$$

Let us calculate the integral $m(\eta)$. We have:
$$
m(\eta)=\int\limits_{0}^{1}\Big(\mu^2-\dfrac{2}{3}\mu\Big)
\Bigg[P\frac{\mu\eta-\eta_1^2}{\eta-\mu}-2\eta_1^2z_0
\frac{\lambda(\eta)}{\eta}\delta(\eta-\mu)\Bigg]d\mu=
$$
$$
=\int\limits_{0}^{1}\Big(\mu^2-\dfrac{2}{3}\mu\Big)(\eta_1^2-\eta
\mu) \dfrac{d\mu}{\mu-\eta}-2\eta_1^2z_0(\eta-\dfrac{2}{3})
\lambda(\eta)\theta_+(\eta).
$$

Here $\theta_+(\eta)$ is the characteristic function for the
interval $0<\eta<1$, i.e.
$$
\theta_+(\eta)= \Big\{
\begin{array}{l}
  1,\quad 0<\eta<1,  \\
  0,\quad -1<\eta<0.
\end{array}
$$
From here we get that
$$
m(\eta)=(\eta^2-\eta_1^2)\Big(\dfrac{1}{6}-\eta\Big)
-(\eta^2-\eta_1^2)\Big(\eta^2-\dfrac{2}{3}\eta\Big)
\ln\Big(1+\dfrac{1}{\eta}\Big)+
$$
$$
+2(\eta^2-\eta_1^2z_0)\Big(\eta-\dfrac{2}{3}\Big),
$$
and when $-1<\eta<0$
$$
m(\eta)=(\eta^2-\eta_1^2)\Big[\dfrac{1}{6}-\eta-\Big(\eta^2-
\dfrac{2}{3}\eta\Big)\ln\Big(1-\dfrac{1}{\eta}\Big)\Big].
$$

The last two formulas can be written as one:
$$
m(\eta)=(\eta^2-\eta_1^2)\Big[\dfrac{1}{6}-\eta-\Big(\eta^2-
\dfrac{2}{3}\eta\Big)f_+(\eta)\Big]+
$$
$$
+2(\eta^2-\eta_1^2z_0)\Big(\eta-\dfrac{2}{3}\Big)\theta_+(\eta),
$$
where
$$
f_+(\eta)=\left\{
\begin{array}{l}
  \ln\dfrac{1+\eta}{\eta},\quad 0<\eta<1,  \\
  \ln\dfrac{1-\eta}{\eta},\quad -1<\eta<0.
\end{array}\right.
$$

We substitute the expansion (6.2) to the condition for the field
(2.7), then this condition can be transformed to the following form:
$$
E_\infty+E_0+\intl_{0}^{1}E(\eta)d\eta=e_0.
\eqno{(7.10)}
$$

Singular integral equation (6.5) is solved completely. The unknown
function $E(\eta)$ is found unambiguously and is determined by the
equality (7.9). The unknown constants can be found from the
equations (7.5), (7.7) and (7.10).

\begin{center}\bf
9. COEFFICIENTS OF THE CONTINUOUS AND DISCRETE SPECTRA
\end{center}

Now for finding of the coefficients of continuous and discrete
spectra we construct system of equations (7.5), (7.10), (7.9)
and (7.7):
$$
z_0A_1=\dfrac{E_0\lambda'(\eta_0)(\eta_1^2-\eta_0^2)-
2E_\infty\lambda_\infty\eta_0}
{(2/3)T(\eta_0)-\lambda_\infty \eta_0},
\eqno{(8.1)}
$$
$$
E_\infty+E_0+\intl_{0}^{1}E(\eta)d\eta=e_0,
\eqno{(8.2)}
$$
or
$$
E_\infty+E_0+\dfrac{1}{2}\intl_{-1}^{1}E(\eta)d\eta=e_0,
\eqno{(8.2)}
$$
$$
E(\eta)=\dfrac{1}{2\eta_1^2z_0\lambda^+(\eta)\lambda^-(\eta)}
\Big[(2E_\infty-1)\lambda_\infty\eta^2+
$$
$$
+z_0A_1\Big(\frac{2}{3}\eta(1+2zT_0(-\eta)
)\Big)],
\eqno{(8.3)}
$$
$$
\frac{1}{36}E_\infty+E_0
m(\eta_0)+\int\limits_{0}^{1}m(\eta)E(\eta)d\eta= $$$$
=-\dfrac{1}{36}z_0A_1\dfrac{1-\alpha_p}{\alpha_p}.
\eqno{(8.4)}
$$

From (8.1) we find the coefficient corresponding to the mode of
Debay:

$$
E_0=\dfrac{(2E_\infty-z_0A_1)\lambda_\infty\eta_0+
\frac{2}{3}z_0A_1T(\eta_0)}{\lambda'(\eta_0)(\eta_1^2-\eta_0^2)}.
\eqno{(8.5)}
$$\smallskip

We substitute the expressions (8.3) and (8.5) into (8.2). From here we
derive:\medskip
$$
E_\infty\Bigg[1-\dfrac{2\lambda_\infty\eta_0}{\lambda'(\eta_0)(\eta_0^2-
\eta_1^2)}+\lambda_\infty J_1\Bigg]+
$$
$$
+z_0A_1\Bigg[-\dfrac{(2/3)T(\eta_0)}
{\lambda'(\eta_0)(\eta_0^2-\eta_1^2)}+\dfrac{\lambda_\infty \eta_0}
{\lambda'(\eta_0)(\eta_0^2-\eta_1^2)}+\dfrac{1}{3}J_2-
\dfrac{\lambda_\infty}{2}J_1\Bigg]=e_0,
$$
where the following designations are introduced:
$$
J_1=\dfrac{1}{2\pi i}\int\limits_{-1}^{1}\Big[\dfrac{1}{\lambda^+(\eta)}-
\dfrac{1}{\lambda^-(\eta)}\Big]\dfrac{\eta
d\eta}{\eta^2-\eta_1^2},
$$
and
$$
J_2=\dfrac{1}{2\pi i}\int\limits_{-1}^{1}\Big[\dfrac{T^+(\eta)}
{\lambda^+(\eta)}-\dfrac{T^-(\eta)}{\lambda^-(\eta)}\Big]
\dfrac{d\eta}{\eta^2-\eta_1^2}.
$$

The integrals $J_1$ and $J_2$ can be calculated analytically with the
help of methods of contour integration. We have

$$
J_1=\Big[\Res_{\infty}+\Res_{-\eta_1}+\Res_{\eta_1}+\Res_{-\eta_0}+
\Res_{\eta_0}\Big]\dfrac{z}{\lambda(z)(z^2-\eta_1^2)}=
$$
$$
=-\dfrac{1}{\lambda_\infty}+\dfrac{1}{\lambda_1}+\dfrac{2\eta_0}
{\lambda'(\eta_0)(\eta_0^2-\eta_1^2)},\quad
\lambda_1=\lambda(\eta_1),
$$
$$
J_2=\Big[\Res_{\infty}+\Res_{-\eta_1}+\Res_{\eta_1}+\Res_{-\eta_0}+
\Res_{\eta_0}\Big]\dfrac{T(z)}{\lambda(z)(z^2-\eta_1^2)}=
$$
$$
=\dfrac{2T(\eta_1)}{2\lambda_1\eta_1}+\dfrac{2T(\eta_0)}{\lambda'(\eta_0)
(\eta_0^2-\eta_1^2)}=\dfrac{1}{2\lambda_1c}+\dfrac{2T(\eta_0)}
{\lambda'(\eta_0)(\eta_0^2-\eta_1^2)},\quad c=\eta_1^2z_0.
$$

We substitute the integral expressions received into the preceding relation. Therefore we obtain 
$$
E_\infty\dfrac{\lambda_\infty}{\lambda_1}+
z_0A_1\Big[\dfrac{1}{6c\lambda_1}-
\dfrac{\lambda_\infty}{2}\Big(\dfrac{1}{\lambda_1}-
\dfrac{1}{\lambda_\infty}\Big)\Big]=e_0,
$$
from where
$$
E_\infty\dfrac{\lambda_\infty}{\lambda_1}+
\dfrac{z_0A_1}{2\lambda_1}\Big[\dfrac{1}{3c}-
(\lambda_\infty-\lambda_1)\Big]=e_0.
$$

Noticing that
$$
\lambda_\infty=\lambda_1+\dfrac{1}{3c},
$$
we find the following relation
$$
E_\infty=e_0\dfrac{\lambda_1}{\lambda_\infty}.
\eqno{(8.6)}
$$

Substituting the deduced expression (8.6) into the equality (8.5), 
we find the following relation connecting $E_0$ and $z_0A_1$
$$
E_0=-2e_0\dfrac{\lambda_1\eta_0}{\lambda'(\eta_0)(\eta_0^2-\eta_1^2)}-
z_0A_1\dfrac{(2/3)T(\eta_0)-\lambda_\infty\eta_0}
{\lambda'(\eta_0)(\eta_0^2-\eta_1^2)},
$$
or
$$
E_0=-2e_0\lambda_1a(\eta_0)-
z_0A_1\Big(\dfrac{2}{3}b(\eta_0)-\lambda_\infty
a(\eta_0)\Big),
\eqno{(8.7)}
$$
where
$$
a(\eta_0)=\dfrac{\eta_0}{\lambda'(\eta_0)(\eta_0^2-\eta_1^2)},\qquad
b(\eta_0)=\dfrac{T(\eta_0)}{\lambda'(\eta_0)(\eta_0^2-\eta_1^2)}.
$$

Let us rewrite the expression for the continuous spectrum 
coefficient in the following form 
$$
E(\eta)=z_0A_1\Big[\dfrac{1}{3c}\dfrac{\eta T_2(\eta)}{\lambda^+(\eta)
\lambda^-(\eta)}-\dfrac{\lambda_\infty \eta^2}{2c \lambda^+(\eta)
\lambda^-(\eta)}\Big]+\dfrac{e_0\lambda_1 \eta^2}{c\lambda^+(\eta)
\lambda^-(\eta)},
\eqno{(8.8)}
$$
where
$$
T_2=1+2\eta T_0(- \eta).
$$

We substitute (8.7), (8.8) and (8.6) into (2.4). We get the equation
$$
\dfrac{e_0\lambda_1}{36\lambda_\infty}-2e_0\lambda_1a(\eta_0)m(\eta_0)-
z_0A_1\Big(\dfrac{2}{3}b(\eta_0)-\lambda_\infty
a(\eta_0)\Big)m(\eta_0)+
$$
$$
+z_0A_1\Big[\dfrac{1}{3}P_2-\dfrac{\lambda_\infty}{2}P_1\Big]+
e_0\lambda_1P_1+z_0A_1\dfrac{1-\alpha_p}{36\alpha_p}=0,
$$
where
$$
P_1=\dfrac{1}{c}\int\limits_{0}^{1}\dfrac{\eta^2 m(\eta)d\eta}
{\lambda^+(\eta)\lambda^-(\eta)}, \qquad
P_2=\dfrac{1}{c}\int\limits_{0}^{1}\dfrac{\eta T_2(\eta) m(\eta)d\eta}
{\lambda^+(\eta)\lambda^-(\eta)}.
$$

From the last equation we find unknown constant $A_1$:
$$
z_0A_1=e_0\lambda_1\dfrac{2a(\eta_0)m(\eta_0)-
\frac{1}{36\lambda_\infty}-P_1}{[\lambda_\infty a(\eta_0)-
\frac{2}{3}b(\eta_0)]m(\eta_0)+\frac{1}{3}P_2-\frac{\Lambda_\infty}{2}P_1+
\frac{1-\alpha_p}{36\alpha_p}}.
\eqno{(8.9)}
$$

Thus, all the coefficients of the expansions (6.1) and (6.2) are
found unambiguously: the coefficient $E_\infty$ is determined
according to (8.6), the coefficient $E_0$ is determined according
to (8.7), the coefficient of the continuous spectrum $E(\eta)$ is
determined according to (8.8), the constant $A_1$ is 
determined from (8.9). Finding of the coefficients of
discrete and continuous spectra of the expansions (6.1) and (6.2)
completes the proving of existence of this expansions. Uniqueness of
the solution in the form of expansions (6.1) and (6.2) can be proved
easily with the use of contraposition method.

\begin{center}\bf
  10. CONCLUSION
\end{center}

In the present work new boundary conditions for the questions of
plasma oscillations in half-space degenerate plasma were proposed. These
boundary conditions are naturally called as specular accommodative
conditions. Such boundary conditions are most adequate for the
problems of {\bf normal} propagation of plasma waves (perpendicular
to the boundary), since accommodation coefficient under such
boundary conditions is {\bf normal} electron momentum accommodation
coefficient.

In the present paper the analytical solution of the problem of
plasma oscillations in half-space degenerate plasma with accommodation of
electron normal momentum, the coefficients of the discrete and
continuous spectra of the problem are found in explicit form.

\end{document}